\documentstyle[11pt,newpasp,twoside,epsf]{article}

\makeatletter
\def\cpr@year{2003}
\def\vol@title{The Neutral ISM in Starburst Galaxies}
\def\vol@author{Susanne Aalto \& Susanne Huttemeister, eds}
\makeatother

\def\eq#1{\begin{equation} {#1} \end{equation}}
\let\mic=\micron
\def\E#1{\hbox{$10^{#1}$}}
\def\sub#1{_{\rm #1}}
\def\x      {\hbox{$\times$}}
\def\about  {\hbox{$\sim$}}
\def\FB    {\hbox{$F\sub{bol}$}}
\def\N     {\hbox{$\cal N$}}
\def\NT    {\hbox{${\cal N}_{\rm T}$}}
\def\Pesc  {\hbox{$P\sub{esc}$}}
\def\S     {\hbox{$S\sub{c\lambda}$}}

\def\tV     {\hbox{$\tau\sub V$}}

\def\Ri     {\hbox{$R\sub{i}$}}
\def\Ro     {\hbox{$R\sub{o}$}}
\def\Mo     {\hbox{$M_\odot$}}
\def\Lo     {\hbox{$L_\odot$}}

\def\deg    {\hbox{$^\circ$}}

\begin{document}
\title{IR emission from AGNs}

\author{Moshe Elitzur}
\affil{Department of Physics and Astronomy, University of Kentucky,
       Lexington, KY 40506, USA}

\author{Maia Nenkova}
\affil{Department of Physics and Astronomy, University of Kentucky,
       Lexington, KY 40506, USA}

\author{\v{Z}eljko Ivezi\'{c}}
\affil{Astrophysical Sciences Department, Princeton
       University, Princeton, NJ 08544, USA}

\begin{abstract}
Unified schemes of active galactic nuclei (AGN) require an obscuring dusty
torus around the central source, giving rise to type 1 line spectrum for
pole-on viewing and type 2 characteristics in edge-on sources. Infrared
radiation at its different wavelengths is the best probe of the dust
distribution, whether the torus orientation is edge-on or pole-on. The observed
IR is in broad agreement with the unified scheme but serious problems remained
in all early modeling efforts. In spite of a general awareness that the dust
must be concentrated in clouds, clumpiness remained the one major ingredient
missing in those radiative transfer studies because of the inherent
difficulties it presents. We have recently developed the formalism to handle
dust clumpiness and our results indicate that its inclusion may resolve the
difficulties encountered by the previous theoretical efforts. We show that
these problems find a natural explanation if the dust is contained in \about\
5--10 clouds along radial rays through the torus. The spectral energy
distributions (SED) of both type 1 and type 2 sources are properly reproduced
from different viewpoints of the same object if the optical depth of each cloud
is $\ga$ 40 at visual wavelengths and the clouds' mean free path increases
roughly in proportion to radial distance.
\end{abstract}

\section{Introduction}
       \label{sec:introduction}

Although there is a bewildering array of AGN classes, a unified scheme has been
emerging steadily (e.g.\ Antonucci 1993, 2002; Wills 1999).  The nuclear
activity is powered by a super\-massive (\about\E6--\E9 \Mo) black hole and its
accretion disk, which extends to \about\ 1 pc. This central engine is
surrounded by a dusty toroidal structure, extending to $\ga$ 100 pc. Much of
the observed diversity is simply the result of viewing this axi\-symmetric
geometry from different angles. The torus provides anisotropic obscuration of
the central region so that sources viewed face-on are recognized as Seyfert 1
galaxies, those observed edge-on are Seyfert 2. The primary evidence for the
torus comes from spectro\-polarimetric observations of type 2 sources, which
reveal hidden type 1 emission via reflection off material situated above the
torus opening. While compelling, this evidence is only indirect in that it
involves obscuration, not direct emission by the torus itself.

\begin{figure}[ht]
\centering \leavevmode
 \epsfxsize=0.7\hsize \epsfclipon \epsfbox{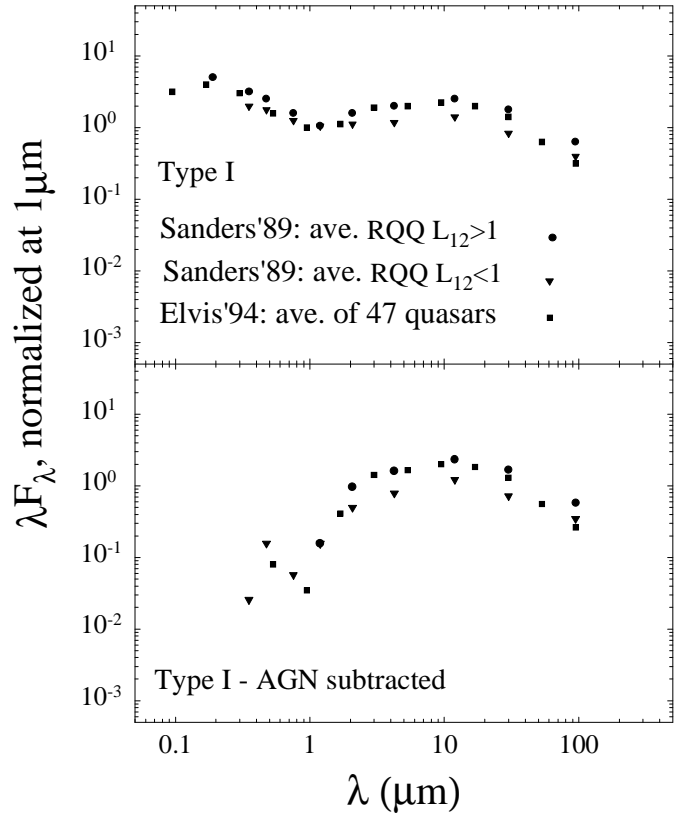}
\caption{SED's of type 1 sources. Top: Average spectra from the indicated
compilations. Bottom: The same SED's after subtracting a power law fit through
the short wavelengths ($\la$ 1\mic). These AGN-subtracted SED's are similar to
those of typical type 2 sources. \label{fig:data1}}
\end{figure}

An obscuring dusty torus should reradiate in the IR the fraction of nuclear
luminosity it absorbs, and the continua from most AGN indeed show significant
IR emission. The top panel of figure \ref{fig:data1} shows composite SED's of
type-1 sources from compilations by Sanders et al (1989) and Elvis et al
(1994). The power law emission at short wavelengths is the expected behavior
from the AGN accretion disk and the excess at $\lambda \ga$ 1\mic\ is
attributed to radiation reprocessed by the dusty torus. Subtraction of the
power law produces the SED's shown in the bottom panel, which are indeed very
similar to those of typical type-2 sources, in accordance with unified schemes.

Pier \& Krolik (1992, 1993) were the first to explore the effects of toroidal
geometry on dust radiative transfer. They noted that the AGN dust must be
concentrated in clouds to protect the grains, but because of the difficulties
in modeling a clumpy medium approximated the density distribution with a
uniform one instead. Still, the directional dependence of their model radiation
did reproduce the gross features of observed SEDs, indicating that the toroidal
geometry captures the essence of IR observations. But major problems remained:
\begin{enumerate}
\item
The models could not generate sufficient emission over the full range of
observed far-IR wavelengths.
\item
The extreme column densities ($\ga$ \E{25} cm$^{-2}$) found in subsequent x-ray
observations greatly exceeded the values allowed by the models. In particular,
such large columns imply extremely deep 10\mic\ absorption feature in type 2
sources, contrary to observations which reveal only moderate depths.
\item
Type 1 sources do not show any evidence of the 10\mic\ silicate feature,
neither in emission nor absorption (e.g., Spoon et al 2003). Such behavior is
reproduced only in a narrow, finely-tuned range of the model parameters.
\item
The huge variation of torus obscuration corresponding to the range of columns
densities found in type 2 x-ray observations (\E{21}--\E{25} cm$^{-2}$) implies
large variation among SED's, incommensurate with the rather modest variation
displayed by the observations.
\end{enumerate}
These problems persisted in all the following studies, which also employed
continuous density distributions (e.g., Granato \& Danese 1994; Efstathiou \&
Rowan-Robinson 1995). Rowan-Robinson (1995) noted that a more realistic model
of the torus would place the dust in clouds and suggested that clumpiness might
alleviate these problems, but was unable to offer a proper treatment of clumps.

\section{Emission from a clumpy medium}

A fundamental difference between clumpy and continuous density distributions is
that radiation can propagate freely between different regions of an optically
thick medium when it is clumpy, but not otherwise. We have recently developed a
detailed formalism for emission from a clumpy medium and applied it to the AGN
obscuring torus (Nenkova, Ivezi\'c \& Elitzur 2002, NIE hereafter, and in
preparation). The building block of our formalism is a single cloud heated both
directly by the AGN and by all other clouds. The direct heating introduces
dependence on the position-angle around the AGN since the dust temperature is
much higher on the cloud's illuminated face than on any other part of its
surface. Therefore the radiation from the cloud depends both on its distance
from the AGN (which controls the highest temperature in the cloud) and the
position angle between the AGN, cloud and observer. We have devised a method
for realistic construction of ``synthetic clump'' source function \S\ that
accounts for this effect and employs an exact solution of radiative transfer
including scattering, absorption and emission by optically thick dust. The
solution is performed with the code DUSTY (Ivezi\'c, Nenkova \& Elitzur 1999).
The absorption and scattering coefficients are those of standard interstellar
dust.

\begin{figure}[ht]
 \centering \leavevmode
 \epsfxsize=0.42\hsize \epsfclipon \epsfbox{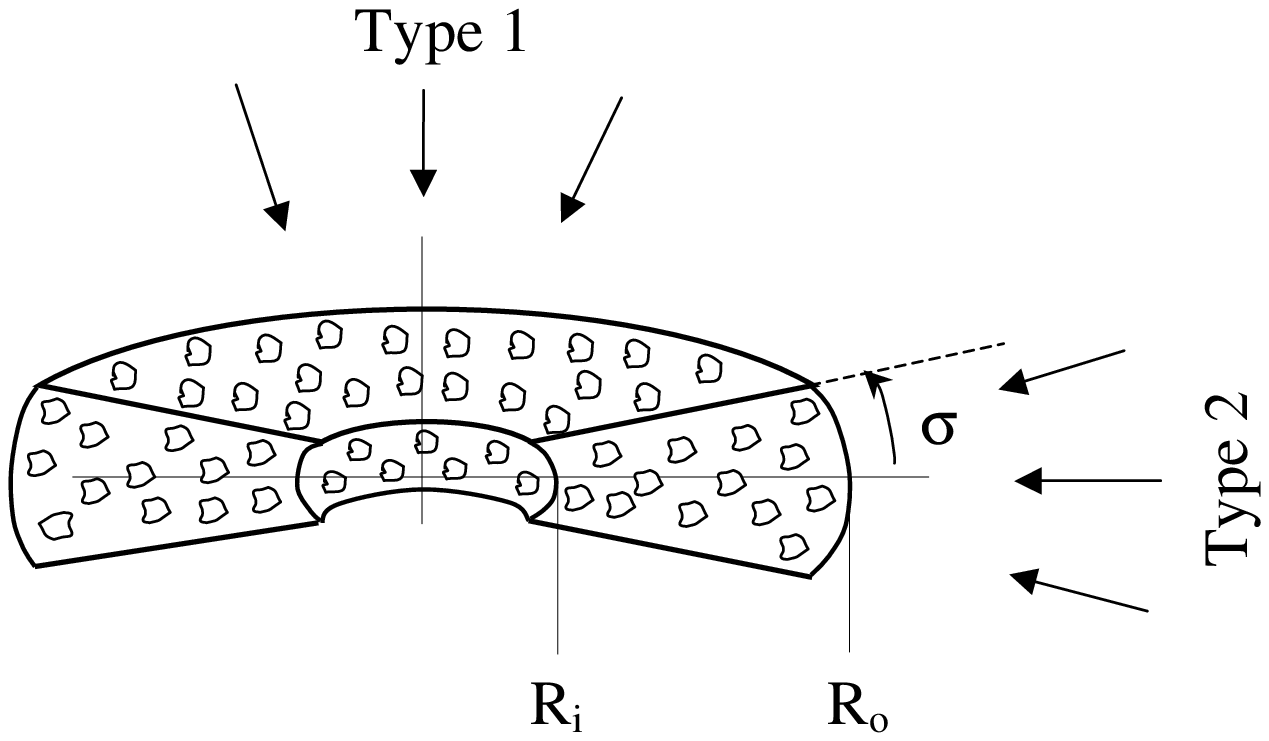} \hfil
 \epsfxsize=0.50\hsize \epsfclipon \epsfbox{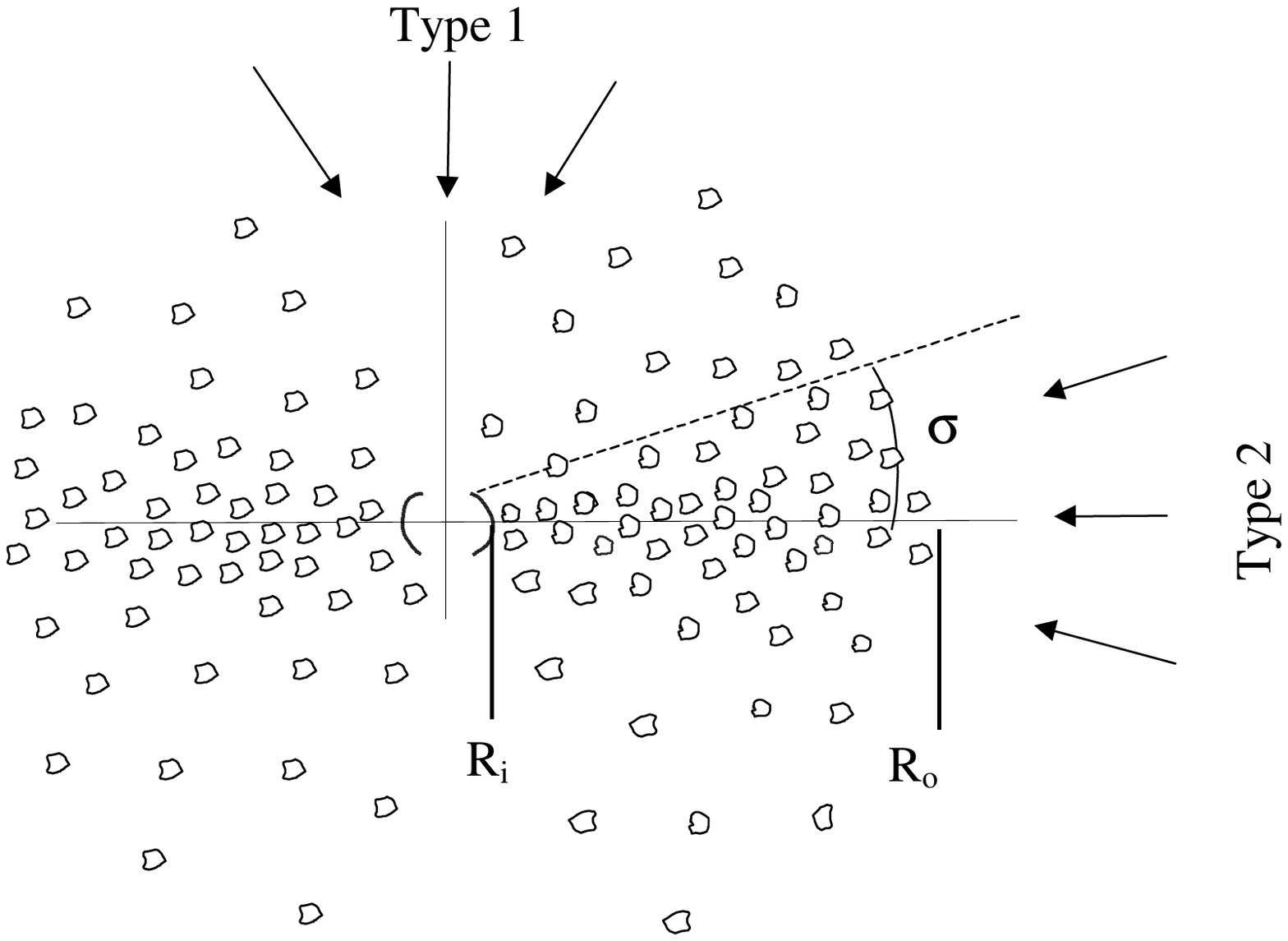} \hfil
\caption{Geometry for modeling AGN clumpy tori between radii \Ri\ and \Ro. The
angular distributions are characterized by a width parameter $\sigma$. The
step-function distribution on the left has a sharp boundary, the one on the
right is smeared, e.g.\ Gaussian. } \label{torus}
\end{figure}

Given these building blocks, we have developed an exact formalism for the
emission when all the dust is in clouds and the cloud distribution follows
Poisson statistics. For simplicity, each cloud has the same optical depth
$\tau_\lambda$, radius $R_c$ and volume $V_c \simeq A_cR_c$ where $A_c$ is the
cloud area. The mean number of clouds encountered in segment $ds$ along a given
path is $d\N(s)= ds/\ell(s)$, where $\ell = (n_cA_c)^{-1}$ is the mean free
path between clouds ($n_c$ is the number density of clouds). The volume filling
factor of the cloud population is then
\eq{
    \phi = n_cV_c = {R_c\over\ell}
}
We adopt as the definition of clumpiness the condition $R_c \ll \ell$, i.e.,
$\phi \ll 1$, small volume filling factor. Under these circumstances, each
cloud can be considered a point source of intensity \S, and the intensity
generated along the segment is $\S d\N$. Denote by \Pesc\ the probability that
this radiation propagate along the rest of the path without absorption by any
other cloud. Natta \& Panagia (1984) show that Poisson statistics for the clump
distribution yields
\eq{\label{eq:Pesc}
  \Pesc = e^{-t_\lambda},            \qquad \hbox{where}\quad
  t_\lambda  = \N(s)( 1 - e^{-\tau_\lambda} )
}
and $\N(s) = \int_sd\N$ is the mean number of clouds along the rest of the
path. The intuitive meaning of this result is straightforward in limit cases.
When $\tau_\lambda < 1$ we simply have $\Pesc = e^{-\tau\sub{tot}}$, the
probability to escape from a continuous medium with overall optical depth
$\tau\sub{tot} = \N\tau_\lambda$. That is, the clumpiness becomes irrelevant.
It is important to note that $\tau\sub{tot}$ can be large, the only requirement
for this limit is that each cloud be optically thin. The opposite limit
$\tau_\lambda \gg 1$ gives $\Pesc = \exp(-\N)$. Even though each clump is
optically thick, a photon can still escape if it avoids all the clumps along
the path.

With this result, the contribution of segment $ds$ to the emerging intensity is
simply $\Pesc\S d\N$ and the overall intensity of a clumpy medium is thus
\eq{\label{eq:IC}
  I^{\rm C}_\lambda = \int e^{-t_\lambda} \S(s)d\N(s),
}
which bears great resemblance to the formal solution of the radiative transfer
equation for continuous medium. This result is rather straightforward and can
be easily generalized to handle configurations in which the clumps have
different properties, and even extended to line radiation. The flux at distance
$D$ is simply $F^{\rm C}_\lambda = \int I^{\rm C}_\lambda dA/D^2$ where $dA$ is
the surface area element perpendicular to the line of sight.

\begin{figure}[ht]
\centering \leavevmode
 \epsfxsize=0.9\hsize \epsfclipon \epsfbox{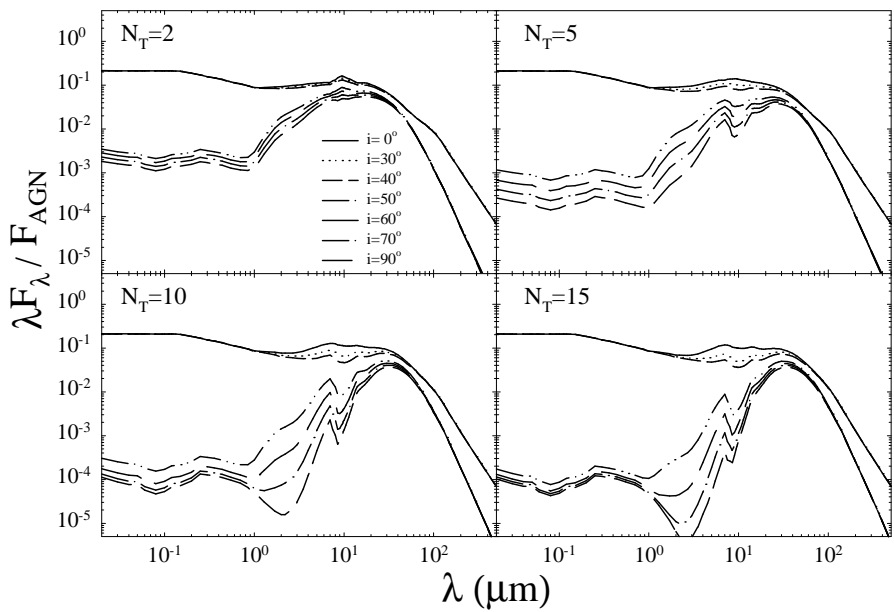}
\caption{Sample model SEDs. The number of clouds \NT\ and the viewing angle
from the torus axis are indicated. In all models, each cloud has optical depth
\tV\ = 60, the cloud radial distribution extends to \Ro\ = 100\Ri\ with the
mean free path increasing in proportion to $r$. The angular distribution is
Gaussian with $\sigma$ = 45\deg. (Nenkova, Ivezi\'c \& Elitzur, in
preparation). \label{fig:SEDs}}
\end{figure}

\begin{figure}[ht]
 \centering \leavevmode
 \epsfxsize=0.5\hsize \epsfclipon \epsfbox{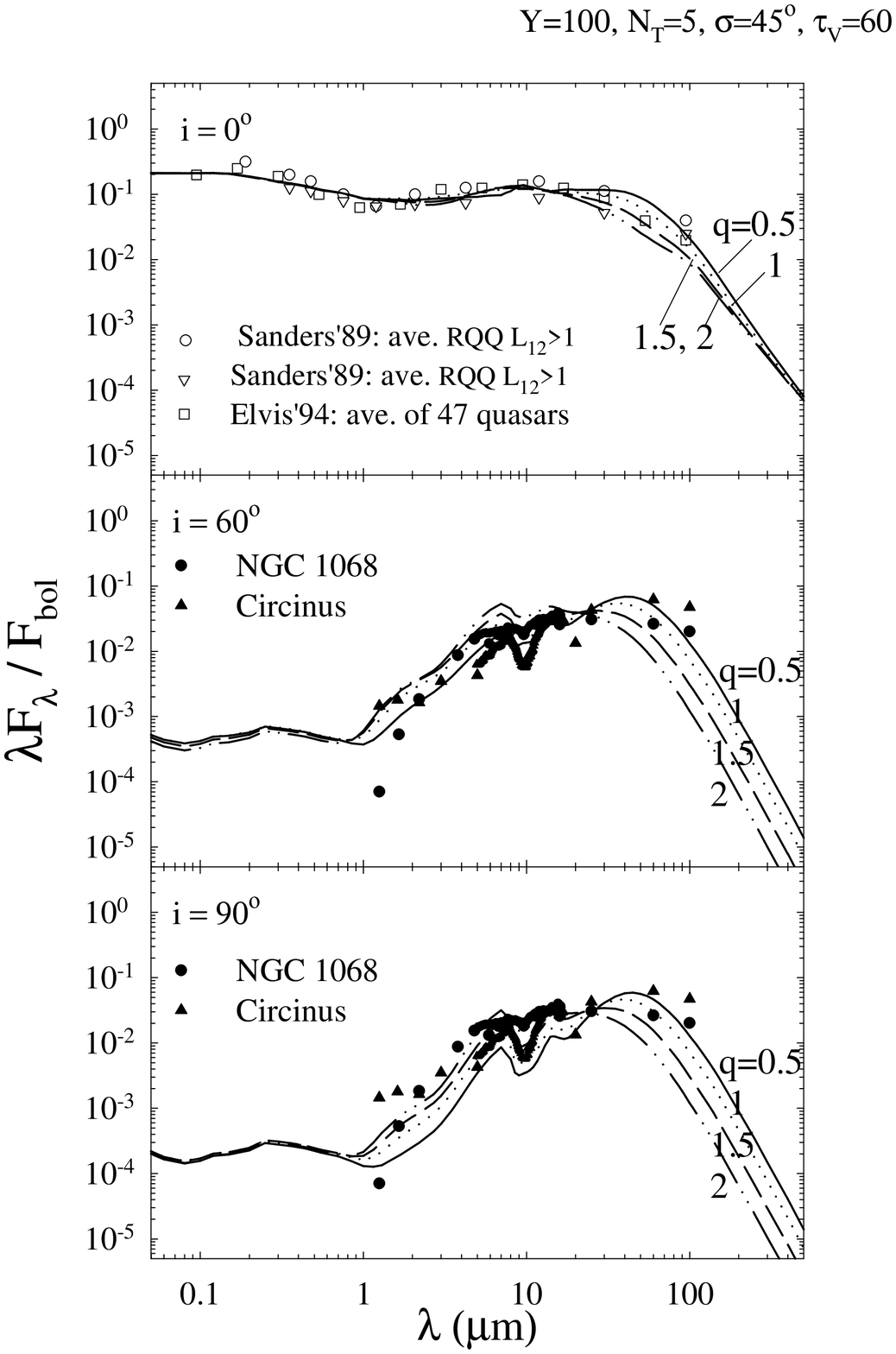}
\caption{Model results for the indicated viewing $i$ from the axis of a
sharp-edge torus with $\sigma$ = 45\deg, \Ro\ = 100\Ri\ and \NT\ = 5 clouds
with \tV\ = 60 each. The clouds' mean free path varies as $r^q$, with the
indicated $q$. The top panel data are type 1 composite spectra. Data in the
bottom panels are for the Seyfert 2 NGC 1068 and Circinus. \label{fig:Data}}
\end{figure}

\section{Application to AGN}

After constructing the clump data base, we applied the formalism to AGN
emission by integrating over toroidal distributions of dusty clouds extending
from inner radius \Ri\ to \Ro, as shown in figure \ref{torus}. In all
calculations \Ri\ occurs where the dust temperature reaches 1400 K, the
temperature above which all grains evaporate. This choice implies \Ri\ =
$1.02\,L_{12}^{1/2}$ pc for an AGN luminosity $L_{12} = L/\E{12} \Lo$. The
cloud radial distribution is described by \NT,  the mean number of clouds along
a ray through the torus equator, and the index $q$ of the power law $r^q$ we
employ for the radial variation of the mean free path between clouds. The
angular distribution is specified by a width parameter $\sigma$. In NIE we
report results for sharp-edge step function angular distributions. We have
since explored continuous distributions and found that Gaussian generally
produces better agreement with observations. The final input is \tV, the
optical depth of each cloud at visual wavelengths.

Figure \ref{fig:SEDs} shows some sample SED's for a set of typical parameters.
Models with only 2 clouds along equatorial rays display a trace of the 10\mic\
feature in emission and thus cannot correspond to the bulk of observed sources.
Similarly, at \NT\ = 15 the near- and mid-IR emission is suppressed too much.
Models with \NT\ = 5--10 clouds produce SED's that closely resemble
observations. It is noteworthy that none of the models displays any of the
severe problems that afflicted those with continuous dust distributions.

According to unified schemes, viewing the same torus from the directions shown
in figure \ref{torus} should give rise to the SEDs observed in type 1 and type
2 sources. Figure \ref{fig:Data} shows sample model outputs and data. Each data
set was scaled for a rough match of the model results without attempting a best
fit. The scaling factor determines the bolometric flux \FB, a quantity
inaccessible for direct measurement for type 2 objects. The panels for $i$ =
0\deg\ and 90\deg\ reproduce results presented in NIE, the $i$ = 60\deg\ panel
is a recent result from work in progress. Data points for type 1 sources show
the average spectra for radio-quiet quasars and Seyfert 1 galaxies, displayed
already in figure \ref{fig:data1}. Because of the high obscuration of the AGN
in type 2 sources, fluxes for their nuclear regions properly extracted out of
the contributions of the host galaxy and starburst regions are scarce. The data
shown are for the prototype Seyfert 2 NGC 1068 and Circinus. As is evident from
the figure, it is possible to reproduce the IR emission from both type 1 and
type 2 sources at different viewing angles of the same clumpy torus, in
accordance with unified schemes. It is interesting that the ``generic'' type 2
inclination angle $i$ = 90\deg\ fails to produce the near IR emission from NGC
1068 and Circinus, but the problem is solved by our new results for $i =
60\deg$. Indeed, Bock et al (2000) conclude from Keck imaging observations that
the inclination is \about\ 65\deg.

The intrinsic luminosity is unobservable in type 2 sources because the bulk of
the  radiation is emitted along directions close to the torus axis. This
persistent problem has especially serious implications for the cosmic x-ray
background (Mushotzky et al 2000). X-ray observations have demonstrated that
there is a large population of heavily obscured AGN, but are unable to measure
the absorbed power with any precision; only IR observations can accomplish this
task. Attempts to determine empirical correlations between the intrinsic x-ray
and mid-IR emission have been made in various observational studies (e.g.,
Alonso-Herrero et al 2001, Krabbe et al 2001). Detailed fits to the SED of
individual sources offer a new, reliable handle on the intrinsic luminosity ---
the overall scaling factor that matches the observed flux density to the
best-fit model SED is simply the AGN bolometric flux \FB.

From comparison with overall properties of observed SED's, and especially the
CfA sample of Seyfert galaxies (Alonso-Herrero et al 2003), we find that the
following set of parameters describes adequately the data:
\begin{itemize}
\item
Gaussian angular distribution with $\sigma = 45\deg \pm 15\deg$
\item
Average number of clouds along equatorial radial rays \NT\ = 5--10
\item
$q$ = 1--2 for the $r^q$ behavior of the mean free path between clouds
\item
Torus outer radius \Ro = (10pc -- 100pc)\x$L_{12}^{1/2}$
\item
Cloud optical depth at visual \tV\ $\ge$ 40
\end{itemize}
The last two points deserve further discussion.

\subsection{Torus dimensions}

\begin{figure}
\centering \leavevmode
 \epsfxsize=0.7\hsize \epsfclipon \epsfbox{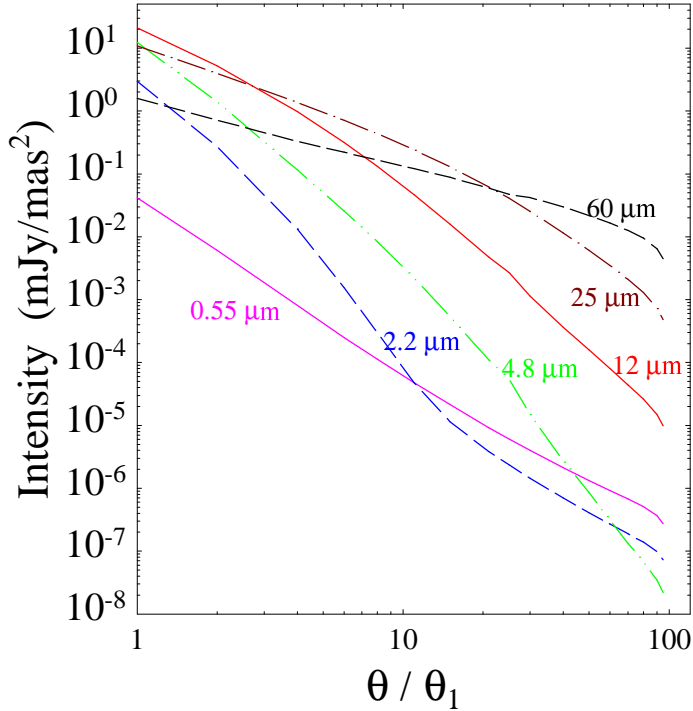}
\caption{Model predictions for intensity profiles at various wavelengths for a
face-on torus with a Gaussian angular distribution of clouds and \Ro\ =
100\,pc\x$L_{12}^{1/2}$. The angular scale $\theta_1$ is 0\farcs02 for a source
with $L_{12} = 1$ at a distance of 10 Mpc.} \label{Fig:Intensity}
\end{figure}

The listed range of acceptable \Ro\ ensures proper reproduction of observed
SEDs. However, pinning down the actual torus size within this range is
surprisingly difficult even with high-resolution imaging observations. Figure
\ref{Fig:Intensity} shows the intensity profiles at various wavelengths for a
torus with \Ro\ = 100\Ri. One striking aspect is the rapid decline of
brightness with distance from the AGN. The torus is the brightest at 12\mic\
but even at this wavelength the FWHM is only \about 2$\theta_1$, and the
angular scale $\theta_1$ is a mere 0\farcs02 for a \E{12}\Lo\ source at a
distance of 10 Mpc. The small dimensions imply that current facilities cannot
resolve the torus even in the most nearby sources. In particular, the \about
0\farcs3 structure visible at the core of NGC 1068 in mid-IR Keck imaging by
Bock et al (2000) reflects instrumental resolution, not the actual torus size.
The torus inner region will be resolvable by the VLTI at its promised angular
resolution of milli-arcsecond at the J-band.

Another problem is that a small torus appears practically indistinguishable
from the inner regions of a larger one. For example, the $\theta \le
30\theta_1$ portions of the plots in figure \ref{Fig:Intensity} are virtually
identical to the full brightness profiles of a torus with \Ro\ = 30\Ri. Because
the near and mid-IR brightness decline by more than two orders of magnitude
already within 10$\theta_1$, the torus outer regions will remain hidden
whenever viewed with instruments with dynamical range of \E2--\E3, and its size
will be underestimated. Thanks to a more moderate brightness decline, longer
wavelengths are more suitable for actual measurements of the torus true size.
Only imaging at \about 60\mic\ by ALMA holds the promise of distinguishing a
30pc torus from a 100pc one in the foreseeable future.

\subsection{Cloud optical depths}

Significantly, the only constraint on the optical depth of individual clouds is
the lower bound \tV\ $\ga$ 40. The model results vary only slightly when \tV\
increases from 40 to 100, and hardly at all during further increase. The reason
is simple. The dependence on \tV\ arises from the probability for photon escape
and the cloud source function, and both factors approach saturation at large
\tV. From eq.\ \ref{eq:Pesc}, $P\sub{esc} = e^{-\cal N}$ whenever $\tau_\lambda
\gg 1$, and at \tV\ $\ga$ 40 this condition is met at all relevant wavelengths.
Similarly, because each cloud is heated from outside, only its surface is
heated significantly when \tV\ is large. Increasing \tV\ further only adds cool
material, thus \S\ saturates for all relevant $\lambda$ (similar to standard
black-body emission). Extending our calculations all the way to \tV\ = 500, we
have verified that increasing \tV\ indeed has no effect on the model results.

This explains at once the surprisingly small variation among AGN SED's in spite
of the huge range of torus columns. From x-ray measurements of 73 Seyfert 2
galaxies Bassani et al (1999) find a large variation in column density, with a
mean of $3\x\E{23}$ cm$^{-2}$ and $\ga$ \E{24} cm$^{-2}$ for as many as a third
of the sources. This mean is comparable with the column density of our torus
for standard dust-to-gas ratio and 5 clouds of \tV\ $\ga$ 40 each. Furthermore,
a natural consequence of our model is the observed similarity of SED among type
2 sources in spite of the large x-ray column variation: since \tV\ is bounded
only from below, the SED remains the same at all $\lambda\ \ga\ 1\mic$ for
arbitrary increase in overall column. In contrast, the x-ray absorption does
vary with \tV\ because the optical depth for Thomson scattering is only \about\
\E{-2}\tV\ so that each cloud remains mostly optically thin. Sources with small
columns may show up in x-ray absorption while selectively excluded from IR
observations because of their weak emission.

\section{Torus Mass}

Risaliti, Maiolino \& Salvati (1999) note that the large column densities
discovered in x-ray absorption imply torus masses in excess of the dynamical
mass, posing a problem for the system stability. For a continuous mass
distribution ($\phi = 1$), the relation between the torus mass $M\sub{torus}$
and the hydrogen overall column density $N\sub{H,tot}$ can be written as
\eq{
  \phi = 1: \qquad
     M\sub{torus} = m\sub{H}N\sub{H,tot}\Omega<r^2>\sub{gas}
  \phantom{XXXXXXX}
}
Here $m\sub{H}$ is the H-mass, $\Omega$ is the solid angle subtended by the
torus at the AGN and $<r^2>\sub{gas}$ stands for the average of $r^2$ weighted
with the gas density distribution. The corresponding relation for a clumpy
torus is
\eq{
  \phi \ll 1: \qquad
    M\sub{torus} = m\sub{H}\NT N\sub{H,C}\Omega<r^2>\sub{clouds}
  \phantom{XXXXXXX}
}
Here $N\sub{H,C}$ is the column density of a single cloud and
$<r^2>\sub{clouds}$ stands for the average of $r^2$ weighted with the cloud
number distribution. Surprisingly, perhaps, the relation for a clumpy torus
does not involve the volume filling factor. Apart from the different average
weighting functions, the two relations give identical masses for the same
overall columns $N\sub{H,tot} = \NT N\sub{H,C}$. However, continuous
distribution usually must be flat to place enough material at large distances
and produce the long wavelength emission. Cloud distributions are steeper in
comparison, because the AGN radiation can reach further thanks to the
clumpiness, resulting in smaller $<r^2>$. With typical parameters, the torus
mass ranges from \E5--\E7\ \Mo, alleviating the mass problem.

\section{Probes of Cloud Distribution}

Our results add strong support for AGN unification schemes. In accordance with
such schemes, the IR emission from both type 1 and type 2 sources is reproduced
at different viewing of the same geometry. Our analysis establishes for the
first time the constraints imposed by the IR emission on cloud distributions,
not just overall column densities. The distribution is described by $q$ and
\NT, individual clouds by their optical depth \tV. No other cloud property was
specified. The cloud size enters only indirectly through the underlying
assumption $R_c \ll \ell$, i.e., $\phi \ll 1$. A reasonable realization, though
not unique, of this assumption is a constant $\phi = 0.1$ throughout the torus.
For a model with $q$ = 1, \NT\ = 5, \tV\ = 100 and $L_{12}$ = 1 this implies
clouds that vary from $R_c$ \about\ 0.1 pc with gas density \about\ 3\x\E5\
cm$^{-3}$ in the torus inner regions to $R_c$ \about\ 10 pc with density
\about\ 3\x\E3\ cm$^{-3}$ at the outer edge, resembling ordinary molecular
clouds.

The IR flux involves integration over the full volume, summing up the
contributions of many lines of sight. By contrast, x-ray absorption measures
the properties of the cloud distribution along a single line of sight. A recent
study by Risaliti, Elvis \& Nicastro (2002) finds x-ray variability on all
available time scales---from months to 20 years, with the latter simply marking
the longest time span currently in the data archives. They conclude that the
x-rays are absorbed by clouds belonging to two populations. The first one
includes clouds at very short distances ($\la$ \E{17} cm) from the AGN. These
are dust free as they are located within the dust sublimation radius. The
second involves clouds at \about\ 5--10 pc, presumably in the dusty torus. The
scenario Risaliti et al propose is similar to a disk-driven hydromagnetic wind
which uplifts, by its ram pressure, and confines, by its magnetic pressure,
dense clouds fragmented from the disk. In the wind inner regions the clouds are
dust free, beyond the sublimation distance they become dusty (Kartje, K\"onigl
\& Elitzur 1999).

Type 1 are distinguished from type 2 sources by the visible AGN contribution
they display. In a clumpy torus, especially with Gaussian edges, the
distinction between the two classes becomes an issue of probability. In spite
of the strong variation with viewing angle, there is always a finite
probability for an unperturbed line of sight to the AGN even in the torus
equatorial plane. Such a source would be classified as type 1 in contrast with
the strict interpretation of unified schemes. The few recorded transitions
between type 1 and 2 may correspond to transits of obscuring clouds across the
line of sight to the AGN.

Our formalism currently handles only the simplest case where all clouds have
identical properties and no intercloud medium. Additional clouds with smaller
\tV\ would make little contribution to IR emission but could absorb the
radiation from the dominant clouds, thus significantly affecting the overall
SED. The same goes for intercloud medium with \tV\ \about 1--5. We plan to
expand our formalism to handle these situations, which may be important for
explaining intermediate (1.x) Seyfert galaxies.

Finally, dust emission provides a snapshot of the torus that carries no
information about its dynamics. Only molecular line observations can provide
information about the kinematics of clouds in the obscuring torus.

\bigskip

\acknowledgements

The support of NSF grant AST-0206149 is gratefully acknowledged. In addition,
M.E thanks Chalmers University for a Jubileum Professor award which supported a
most enjoyable visit to Onsala Space Observatory.

\end{document}